\documentstyle[tighten,twocolumn,aps,psfig]{revtex}
\draft
\begin{document}
\title{ Why the high lying glueball does not mix
with the neighbouring  $f_0$.}
\author{ L. Ya. Glozman}
\address{  Institute for Theoretical
Physics, University of Graz, Universit\"atsplatz 5, A-8010
Graz, Austria\footnote{e-mail: leonid.glozman@uni-graz.at}}

\twocolumn[
    \begin{@twocolumnfalse}
      \maketitle
      \widetext

\begin{abstract} 
Chiral symmetry restoration in high-lying hadron spectra
implies that hadrons which belong to different irreducible
representations of the parity-chiral group cannot mix. 
This explains why 
the $f_0(2102 \pm 13)$, which was suggested to be a glueball,
and hence must belong to the scalar (0,0) representation of
the chiral group, cannot mix with the neighbouring $f_0(2040 \pm 38)$,
which was interpreted as a $ n\bar n$ state, and that belongs to the
$(1/2,1/2)$ representation of the chiral group. If confirmed,
then we have an access to a "true" glueball of QCD.
\end{abstract}
\bigskip
\bigskip
\end{@twocolumnfalse}
  ]
{
    \renewcommand{\thefootnote}%
      {\fnsymbol{footnote}}
    \footnotetext[1]{e-mail address: leonid.glozman@uni-graz.at}
    }
\narrowtext

\section{Introduction}

The puzzle of the $0^{++}$ $f_0$ resonances has
attracted a significant attention for the last
decade (for recent reviews and references see
\cite{PDG,AN,CL}). The reason is that there
have been discovered more $f_0$ mesons
than can be accomodated by the quark model. This also
fits with the expectations that there should be a
glueball state with the same quantum numbers around the
1.5 GeV region . And indeed, an analysis of
numerous available experimental data suggests that
$f_0(1370)$ is mostly a $ n\bar n = 
\frac{u\bar u + d\bar d}{\sqrt 2}$ state, $f_0(1710)$
is mostly a $s\bar s$ state and $f_0(1500)$ is dominantly
 a glueball \cite{AM}. Other alternatives have also been discussed
\cite{O}. The decay modes of these mesons  suggest
that the physical mesons  above are  some mixtures
of the pure quark-antiquark and glueball states.\\

Recently there have appeared results of the partial
wave analysis of the mesonic resonances obtained in
the $p \bar p$ annihilation at LEAR  in the region
1.8 - 2.4 GeV \cite{BUGG1,BUGG2,BUGG3}. In particular,
 four high-lying $f_0$ have been reported:

$$ f_0(1770 \pm 12),f_0(2040 \pm 38),f_0(2102 \pm 13),
f_0(2337 \pm 14). $$ 

\noindent
It has been suggested that $f_0(2102 \pm 13)$ should be a
glueball, while $f_0(1770 \pm 12)$, $f_0(2040 \pm 38)$
 and  $f_0(2337 \pm 14)$
 are  $n \bar n$ states. Motivation for
such an interpretation was the following: (i) all these
states are observed in $p \bar p$, hence according to
OZI rule they cannot be $s \bar s$ states; (ii) 
$f_0(2102 \pm 13)$ decays more strongly to $\eta \eta$
than to $\pi \pi$. It requires a large mixing angle,
in contrast to the other three $f_0$ states. It can be
interpreted naturally only if  $f_0(2102 \pm 13)$ is not
a $n \bar n$ state, but a glueball. \\

If $f_0(2102 \pm 13)$ is a glueball, a question then arises
 as  why it  does not  mix with the neighbouring
broad $f_0(2040 \pm 38)$, which is supposed to be a $n \bar n$
state? {\it "When a particle is observed experimentally
it does not come labeled "quark-antiquark" or "glueball",
nor is there any strict objective criterion to distinguish
them. The whole distinction is tied up with the naive
quark model, which has no firm basis in quantum field theory,
and ignores the inevitability of mixing." }\cite{WIL}.\\

The aim of the present short note is to offer a natural
explanation of absence of a strong mixing. It has been
argued recently that high in the hadron spectrum
the spontaneously broken chiral symmetry of QCD is effectively
restored \cite{G1,CG1,CG2,G2,G3}\footnote {This phenomenon has been
refered  to\cite{G4} as {\it chiral symmetry restoration of the second kind}
in order to distinguish it from the familiar phenomenon
of chiral symmetry restoration in the QCD vacuum at high
temperature and/or density.}. A possible fundamental physical
ground is that, in hadrons with large $n$ (radial quantum number)
or large $L$ the semiclassical approximation must be valid
and hence the effects of quantum fluctuations of the quark and gluon fields
must be suppressed \cite{G5}.
If chiral symmetry is
approximately restored, then the index of the representation of
the chiral group, to which  the given hadron belongs, becomes
a good quantum number. Hence a mixing of  hadrons that
belong to different representations is forbidden, {\it
even though all other quantum numbers are the same and
hadrons are close in energy.} This then explains the absence
of a strong mixing between $f_0(2040 \pm 38)$ state, which
is a member of the (1/2, 1/2) representation of the chiral
group \cite{G2}, with a glueball, which belongs to a scalar
representation (0,0).\\

\section{ The evidence and theoretical justification of
chiral symmetry restoration in high-lying hadrons}

We overview first a QCD-based justification
of chiral symmetry restoration in high-lying spectra \cite{CG1,CG2}.
Consider two local currents (interpolating fields) $J_1(x)$ and 
$J_2(x)$ which are connected by chiral transformation,
$J_1(x) = U J_2(x) U^\dagger$, where $U \in SU(2)_L \times SU(2)_R$.
These currents, when act on the QCD vacuum $|0\rangle$, create
hadron states with  quantum numbers "1" and "2", respectively.
All these hadrons are the intermediate states in the two-point
correlators

\begin{equation}
\Pi_{J_\alpha}(q) = i \int d^4x~ e^{-iqx}
\langle 0 | T \left \{ J_\alpha(x) J_\alpha(0)\right \} | 0 \rangle,
\label{corr}
\end{equation}

\noindent
where all possible Lorentz indices (which are specific for
a given interpolating field) have been omitted, for simplicity.
At large space-like momenta $Q^2 = -q^2 > 0$ the correlator
can be adequately represented by the operator product
expansion, where all nonperturbative effects reside in
different condensates \cite{SVZ}. The only effect that
spontaneous breaking of chiral symmetry can have on the
correlator is via the quark condensate of the vacuum,
$\langle q \bar q \rangle$, and higher dimensional
condensates that are not invariant under chiral transformation $U$.
However, the contributions of all these condensates are suppressed
by inverse powers of momenta $Q^2$.  This shows that
at large space-like momenta the correlation function
becomes chirally symmetric. In other words

\begin{equation}
\Pi_{J_1}(Q) \rightarrow  \Pi_{J_2}(Q) ~~~ at ~~~Q^2 \rightarrow \infty.
\label{c0}
\end{equation}

\noindent
The dispersion relation provides a connection between the
space-like and time-like domains for the Lorentz scalar
(or pseudoscalar) parts of the correlator. In particular,
the large  $Q^2$ correlator is completely dominated by the
large $s$ spectral density $\rho(s)$, which is an observable.
Hence the large $s$ spectral density should be insensitive
to the chiral symmetry breaking in the vacuum and must
satisfy 

\begin{equation}
\rho_1(s) \rightarrow  \rho_2(s) ~~~ at ~~~s \rightarrow \infty.
\label{c1}
\end{equation}

\noindent
This is in contrast to the low $s$ spectral densities 
$\rho_1(s)$ and $\rho_2(s)$, which are very different
because of the chiral symmetry breaking in the vacuum.
This manifests a smooth chiral symmetry restoration
from the low-lying spectrum to the high-lying spectrum
(chiral symmetry restoration of the second kind).
\footnote{A theoretical
expectation that chiral symmetry must be restored high in
the spectra is supported by the recent data on the
difference between the vector and axial vector spectral
densities. This difference has been extracted from the weak
decays of the $\tau$-lepton  by the ALEPH  and OPAL
collaborations \cite{ALEPH,OPAL}. The nonzero
difference is entirely from the spontaneous breaking of
chiral symmetry. It is well seen from the results that
while the difference is large at the masses of $\rho(770)$
and $a_1(1260)$, it becomes  very strongly reduced towards
$m = \sqrt s \sim 1.7$ GeV. This is also seen from 
$e^+e^- \rightarrow hadrons$, where starting approximately
from the same energy the spectral density oscillates around
perturbative QCD prediction \cite{EE}. Similarly, recent data
of JLAB on inclusive electroproduction of baryonic resonances
in the mass region $3.1 \leq M^2 \leq 3.9$ GeV$^2$ are 
perfectly dual to deep inelastic data \cite{N}.}\\

Since the inclusive data indicate that the quark-hadron duality
picture starts to work  in the resonance region,
we have to anticipate in this region
a nontrivial implication of chiral symmetry. Indeed, if
 chiral symmetry restoration happens in the regime where
the spectrum is still quasidiscrete (i.e. it is dominated
by resonances and the successive
resonances with the given spin are well separated), then
these resonances must fill out representations of the parity-chiral
group. There are evidences both in baryon \cite{G1,CG1,CG2}
and in meson spectra \cite{G2,G3} that  light hadrons
above $m \sim 1.7$ GeV fill out representations of
the parity-chiral group, which are manifest as parity doublets
or higher chiral representations. Nevertheless, a systematic
experimental exploration of the high-lying hadrons is required 
in order to make definitive statements.\\

While the asymptotic prediction (\ref{c1}) is rather robust 
(it is based only on the asymptotic freedom of QCD in the deep
space-like domain and on the analyticity of the two-point
correlator; earliest application of asymptotic freedom  to
$e^+e^- \rightarrow hadrons$  is ref. \cite{AG}), 
it does not tell us which physics could be associated
with the chiral symmetry restoration
in the isolated hadron. This question has been addressed
in ref. \cite{G5}. One of the possibilities is that the
chiral symmetry breaking (i.e. dynamical quark mass generation)
is due to  quantum fluctuations of the quark and gluon fields.
This can be seen from the fact that the chiral symmetry breaking
can be formulated via the Schwinger-Dyson equation. For the
present context it is not important at all which specific gluonic
interactions are the most important ones in the kernel
of the Schwinger-Dyson equation, instantons, gluon exchanges
or anything else. If the effects of quantum fluctuations of the quark and
gluon fields are suppressed, then the dynamical mass of quarks
must vanish. At large $n$ (radial quantum number) or at large
angular momentum $L$ we know that in quantum systems the 
effects of quantum
fluctuations become indeed suppressed and the semiclassical approximation
(WKB) must work.\footnote{For example, the Lamb shift in the hydrogen
atom is entirely due to the quantum fluctuations of the electromagnetic
and electron fields, which are the vertex correction, electron self-energy
and the vacuum polarization diagrams. The Lamb shift vanishes
very fast with $n$, $\sim 1/n^3$, and also very fast with $L$.} 
Physically, this approximation applies in these cases because the
de Broglie wavelength of the valence quarks in the hadron is
small in comparison with the size of the hadron. If so the chiral
symmetry must be effectively restored in high-lying hadrons.  So a very natural picture for highly excited
hadrons is a string with bare quarks of definite chirality at
the end-points \cite{G3}.

\section {Chiral quantum number as a good quantum number}

In the low-lying hadrons, where the effects of  chiral
symmetry breaking in the vacuum 
$SU(2)_L \times SU(2)_R \rightarrow SU(2)_I$ are strong,
only isospin is a good quantum number among those that
are potentially supplied by the chiral group. In the
regime where  chiral symmetry is (approximately)
restored a new good quantum number appears, that characterizes
a parity-chiral multiplet. This quantum number is an index
of the irreducible representation of the parity-chiral 
group\footnote{The irreducible representations of 
$SU(2)_L\times SU(2)_R$
can be labeled as $(I_L, I_R)$ with $I_L$ and $I_R$
being the isospins of the left and right subgroups. However,
generally the states that belong to the given irreducible representation
of the chiral group cannot be ascribed a definite parity
because under parity transformation the left-handed quarks
transform into the right-handed ones (and vice versa). Therefore under
a parity operation the irreducible representation $(I_L, I_R)$ transforms
into $(I_R, I_L)$. Hence, in general, the state (or current)
 of definite parity can
be constructed as a direct sum of two irreducible representations
$(I_L, I_R) \oplus (I_R, I_L)$, which is an irreducible
representation of the parity-chiral group \cite{CG2}.}.
For example, the vector ($\rho$) and axial vector ($a_1$) mesons,
in the chirally restored regime become chiral partners and
fill out in pairs $(0,1) \oplus (1,0)$ irreducible representations
of the parity-chiral group. Their valence content is given as \cite{G3}

\begin{equation}
\rho: ~~~  \frac{1}{\sqrt 2}\left( \bar R {\vec \tau}
\gamma^\mu R + \bar L {\vec \tau} \gamma^\mu L \right);
\label{rho}
\end{equation}

\begin{equation}
a_1: ~~~  \frac{1}{\sqrt 2}\left( \bar R {\vec \tau}
\gamma^\mu R - \bar L {\vec \tau} \gamma^\mu L\right).
\label{a1}
\end{equation}

The valence composition of $n \bar n$ $f_0$ and $\pi$ mesons
in the chirally restored regime is 

\begin{equation}
f_0: ~~~  \frac{1}{\sqrt 2}\left( \bar R 
L + \bar L  R\right);
\label{f0}
\end{equation}

\begin{equation}
\pi: ~~~  \frac{1}{\sqrt 2}\left( \bar R {\vec \tau}
 L - \bar L {\vec \tau}  R\right).
\label{pion}
\end{equation}

\noindent
Both these mesons fill out in pairs the (1/2,1/2) representations
of the parity-chiral group \cite{G2}. The
experimental data in the range 1.8 - 2.4 GeV are summarized
in the Table.

\begin{center}
\begin{tabular}{|llllll|} \hline
Meson & ~I~ & $~J^P~$ & Mass (MeV) & Width (MeV) & Reference\\ \hline
$f_0$ & ~0~ & $~0^+~$  & $1770 \pm 12$ &  $220 \pm 40$ & \cite{BUGG1}\\
$f_0$ & ~0~ & $~0^+~$  & $2040 \pm 38 $ &  $405 \pm 40$ & \cite{BUGG2} \\
$f_0$ & ~0~ & $~0^+~$  & $2102 \pm 13$  &  $211 \pm 29$ & \cite{BUGG2} \\
$f_0$ & ~0~ & $~0^+~$  & $2337 \pm 14$  &  $217 \pm 33$ & \cite{BUGG2} \\
$\pi$ & ~1~ & $~0^-~$  & $1801 \pm 13$   &  $210 \pm 15$ & \cite{PDG} \\
$\pi$ & ~1~ & $~0^-~$  & $2070 \pm 35$   &  $310^{+100}_{-50}$ & \cite{BUGG3}\\
$\pi$ & ~1~ & $~0^-~$  & $2360 \pm 25$   &  $300^{+100}_{-50}$ & \cite{BUGG3}\\
\hline
\end{tabular}
\end{center}

It is well seen that while the chiral symmetry is strongly
broken low in the spectrum, the high-lying 
$n \bar n$ $f_0$ and $\pi$ mesons indeed form chiral pairs (see also Fig. 1):

\begin{equation}
\pi(1300 \pm 100) - f_0(1370_{-170}^{+130}),
\end{equation}

\begin{equation}
\pi(1801 \pm 13) - f_0(1770 \pm 12),
\end{equation}

\begin{equation}
\pi(2070 \pm 35) - f_0(2040 \pm 38 ),
\end{equation}

\begin{equation}
\pi(2360 \pm 25) - f_0(2337 \pm 14).
\end{equation}
\\

A true glueball (G) has no valence quark content and
hence must belong to the scalar representation of the
parity-chiral group, $G \sim (0,0)$.\\

\begin{figure}
\hspace*{-0.5cm}
\centerline{
\psfig{file=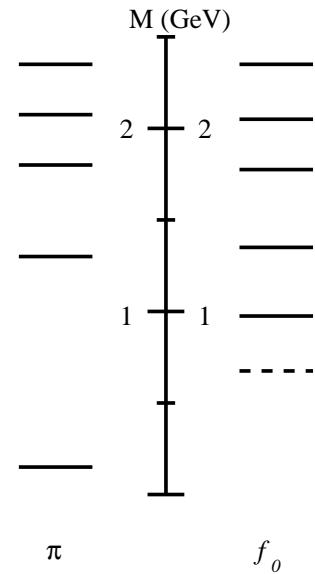,angle=-90,width=0.6\textwidth}
}
\caption{Pion and $n \bar n$ $f_0$ spectra. The three highest
states in both pion and $f_0$ spectra are taken from \protect\cite{BUGG1,BUGG2,BUGG3}.
Since  these $f_0$ states are obtained in $p \bar p$ and   they
decay predominantly  into $\pi \pi$ channel, they are considered in 
\protect\cite{BUGG1,BUGG2,BUGG3} as  $n \bar n$ states.}
\end{figure}

If chiral symmetry is a good symmetry then the index of the
corresponding irreducible representation of the 
parity-chiral group becomes a good quantum number. Hence
quantum mechanics forbids a mixing of the states that
belong to different representations, even though all other
quantum numbers coincide and states are close in energy.
Since the true glueball by definition belongs to
the $(0,0)$ representation and the approximately degenerate
$f_0$ and $\pi$ mesons form the $(1/2,1/2)$ representation,
they cannot be mixed. This explains why
 $f_0(2102)$, which is
presumably a glueball and has no partner in the pion spectrum,
and which is very close to $f_0(2040)$, that is
a $n \bar n$ state, does not mix with the latter. 
If confirmed by a detailed study of
decay modes, it would be better to rename $f_0(2102)$
as $G(2102)$.\\

Clearly, the chiral restoration is not exact at masses of
2 GeV, so a small amount of mixing is still possible.
Hence, it is a very important experimental task to
study in detail decay modes and mixings of these high-lying
states. The prize for these efforts would be that we
identify and study a "true" glueball.\\

We can also look at the same problem  from the  other perspective.
If a detailed study of decay modes confirmes that $f_0(2102)$
is a glueball and that $f_0(2040)$ is a $n \bar n$ state,
then it would be an independent confirmation of chiral
symmetry restoration.\\

As a conclusion, chiral symmetry restoration high in the 
hadron spectra provides a natural basis for the absence of
a strong mixing between those scalar mesons that are
chiral partners of pions, and those scalar mesons, that
represent a glueball. Then we have an opportunity to study a
"true" glueball of QCD.\\

{\bf Acknowledgements}

\medskip
The work was supported by the FWF project P14806-TPH
of the Austrian Science Fund.


\bigskip



\begin{references}
\bibitem{PDG} Particle Data Group, D. Groom et al, Eur. Phys. J
{\bf C15}, 1 (2000).
\bibitem{AN} V. V. Anisovich, Physics-Uspekhi, {\bf 41}, 419 (1998).
\bibitem{CL} F. E. Close and N. A. T\"ornqvist, hep-ph/0204205.
\bibitem{AM} C. Amsler, Phys. Lett. {\bf B451},22 (2002).
\bibitem{O} N. A. T\"ornqvist, Z. Phys. {\bf C68}, 647 (1995);
M. Boglione and M. R. Pennington, Phys. Rev. Lett. {\bf 79}, 1998 (1997);
A. V. Anisovich, V. V. Anisovich, A. V. Sarantsev,  
Z. Phys. {\bf A357}, 123 (1997); P. Minkowski and W. Ochs, Eur. Phys. J,
{\bf C9}, 283 (1999); D. Weingarten, Nucl. Phys. Proc. Suppl.,
{\bf 73}, 249 (1999); V.V. Anisovich, hep-ph/0208123.
\bibitem{BUGG1} A. V. Anisovich et al, Phys. Lett. {\bf B449}, 154 (1999);
     BES Collaboration, Phys. Lett. {\bf B472}, 207 (2000).
\bibitem{BUGG2} A. V. Anisovich et al, Phys. Lett. {\bf B491}, 47 (2000).
\bibitem{BUGG3} A. V. Anisovich et al, Phys. Lett. {\bf B517}, 261 (2001).
\bibitem{WIL} F. Wilczek, hep-lat/0212041.
\bibitem{G1} L. Ya. Glozman, Phys. Lett. {\bf B475}, 329 (2000).
\bibitem{CG1} T. D. Cohen and L. Ya. Glozman, Phys. Rev. {\bf D65}, 016006 (2002).
\bibitem{CG2} T. D. Cohen and L. Ya. Glozman,  Int. J. Mod. Phys. 
{\bf A17}, 1327 (2002).
\bibitem{G2} L. Ya. Glozman, Phys. Lett. {\bf B539}, 257 (2002).
\bibitem{G3} L. Ya. Glozman, Phys. Lett. {\bf B541}, 115 (2002).
\bibitem{G4} L. Ya. Glozman, hep-ph/0210216.
\bibitem{G5} L. Ya. Glozman, hep-ph/0304087.
\bibitem{SVZ} M. A. Shifman, A. I. Vainstein, V. I. Zakharov,
Nucl. Phys. {\bf B147}, 385 (1979).
\bibitem{ALEPH} ALEPH collaboration: R. Barate et al, Eur. Phys. J.
{\bf C4}, 409 (1998).
\bibitem{OPAL} OPAL collaboration: K. Ackerstaff et al, Eur. Phys. J.,
{\bf C7}, 571 (1999).
\bibitem{EE} M. Davier, S. Eidelman, A. H\"ocker, and Z. Zhang,
hep-ph/0208177.
\bibitem{N} I. Niculesku et al, Phys. Rev. Lett. {\bf 85}, 1186  (2000).
\bibitem{AG} T. Appelquist and H. Georgi, Phys. Rev. {\bf D8}, 4000 (1973). 
\end{references}
\end{document}